\documentclass[sigconf, nonacm]{acmart}
\AtBeginDocument{%
  }

\setcopyright{acmlicensed}
\copyrightyear{2026}
\acmYear{2026}
\setcopyright{cc}
\setcctype{by}
\acmConference[CHASE'26]{International Workshop on Cooperative and Human Aspects of Software }{April 12--18, 2026}{Rio de Janeiro, Brazil}
\acmBooktitle{International Workshop on Cooperative and Human Aspects of Software (CHASE'26), April 12--18, 2026, Rio de Janeiro, Brazil}
\acmPrice{}
\acmDOI{10.1145/3794860.3794903}
\acmISBN{979-8-4007-2484-8/2026/04}

\acmISBN{978-1-4503-XXXX-X/2018/06}

\usepackage{multirow}
\usepackage{subcaption}
\usepackage{caption}
\usepackage{soul}
\usepackage{graphicx}     % Para incluir imagens

\begin{document}

\title[A Quasi-Experimental Evaluation of Coaching to Mitigate the IP in Early-Career Software Engineers]{A Quasi-Experimental Evaluation of Coaching to Mitigate the Impostor Phenomenon in Early-Career Software Engineers}


\author{Paloma Guenes}
\email{pguenes@inf.puc-rio.br}
\orcid{0009-0004-8080-1760}
\affiliation{%
  \institution{Pontifical Catholic University of Rio de Janeiro}
  \city{Rio de Janeiro}
  \country{Brasil}
}
\affiliation{%
  \institution{University of Bari}
  \city{Bari}
  \country{Italy}
}

\author{Joan Leite}
\email{joan_leite@hotmail.com}
\orcid{0000-0003-1707-8093}
\affiliation{%
  \institution{Pontifical Catholic University of Rio de Janeiro}
  \city{Rio de Janeiro}
  \country{Brasil}
}

\author{Rafael Tomaz}
\email{rafaelstomaz@gmail.com}
\orcid{0009-0009-1435-4372}
\affiliation{%
  \institution{Pontifical Catholic University of Rio de Janeiro}
  \city{Rio de Janeiro}
  \country{Brasil}
}

\author{Allysson Allex Araújo}
\email{allysson.araujo@ufca.edu.br}
\orcid{0000-0003-2108-2335}
\affiliation{%
  \institution{Federal University of Cariri}
  \city{Juazeiro do Norte}
  \country{Brasil}
}

\author{Jean Natividade}
\email{jeannatividade@gmail.com}
\orcid{0000-0002-3264-9352}
\affiliation{%
  \institution{Pontifical Catholic University of Rio de Janeiro}
  \city{Rio de Janeiro}
  \country{Brasil}
}

\author{Maria Teresa Baldassarre}
\email{mariateresa.baldassarre@uniba.it}
\orcid{0000-0001-8589-2850}
\affiliation{%
  \institution{University of Bari}
  \city{Bari}
  \country{Italy}
}

\author{Marcos Kalinowski}
\email{kalinowski@inf.puc-rio.br}
\orcid{0000-0003-1445-3425}
\affiliation{%
  \institution{Pontifical Catholic University of Rio de Janeiro}
  \city{Rio de Janeiro}
  \country{Brasil}
}

\renewcommand{\shortauthors}{Guenes et al.}

\begin{abstract}
% [Context] 
[Context] The Impostor Phenomenon (IP), the persistent belief of being a fraud despite evident competence, is common in Software Engineering (SE), where high expectations for expertise and innovation prevail. Although coaching and similar interventions are proposed to mitigate IP, empirical evidence in SE remains underexplored. [Objective] This study examines the impact of a structured group coaching intervention on reducing IP feelings among early-career software engineers. [Method] We conducted a quasi-experiment with 20 participants distributed across two project teams using a wait-list control design, complemented by non-participant observation. The treatment group received a three-session coaching intervention, while the control group received it after an observation phase. IP was assessed using the Clance Impostor Phenomenon Scale (CIPS), alongside evaluated measures of well-being (WHO-5), life satisfaction (SWLS), and affect (PANAS). [Results] The coaching resulted in modest reductions in CIPS scores, whereas the control group also improved during the observation phase, suggesting that contextual and temporal factors may have exerted stronger influence than the formal intervention. [Conclusion] These results suggest that coaching may support reflection and awareness related to IP, yet other contextual aspects of team collaboration and project work might also contribute to these changes. This study offers a novel empirical step toward understanding how structured IP interventions operate within SE environments.

%This study provides one of the first empirical evaluations of an IP-targeted intervention in a SE context. We conclude that IP among early-career software engineers should be understood as a malleable state, directly shaped by workplace context. Hence, we highlight that contextual dynamics may be more influential on participants' IP and related psychological states than the formal coaching. 

%This study provides one of the first empirical evaluations of an IP-targeted intervention in a SE context. It reinforces both the potential of coaching interventions and the importance of contextual dynamics in improving software engineers’ professional well-being.

\end{abstract}






\maketitle

\section{Introduction}

The Impostor Phenomenon (IP) refers to a persistent self-perception of intellectual fraudulence despite clear evidence of competence \cite{clance1978imposter}. Individuals experiencing IP tend to attribute their achievements to luck or external factors rather than ability, accompanied by a chronic fear of being exposed as a fraud. In Software Engineering (SE), such experiences are particularly salient \cite{Guenes24}. The field’s continuous technological change, peer evaluation, and high expectations for innovation create fertile ground for self-comparison and perceived inadequacy \cite{maji2021they, oliveira2024navigating}.

%The Impostor Phenomenon (IP) is a psychological pattern wherein individuals doubt their accomplishments and harbor a persistent fear of being exposed as a fraud, despite evidence of their competence \cite{clance1978imposter}. This phenomenon is not uncommon in high-pressure and rapidly evolving fields such as software engineering \cite{maji2021they, Guenes24, Guenes25, oliveira2024navigating}, where professionals are often expected to exhibit high levels of expertise and innovation. In fact, the consequences of suffering from IP may contribute to mental health disorders, such as depression and burnout \cite{sakulku2011impostor}, and it is also present in software engineering \cite{tulili2023burnout}.

Recent surveys report that more than half of SE professionals experience frequent or intense impostor feelings, often associated with reduced perceived productivity \cite{Guenes24} and well-being \cite{Guenes25}. Underrepresented groups (for instance women and minority professionals) report even higher prevalence and stronger correlations with poor well-being \cite{oliveira2024navigating, Guenes25}. Beyond transient self-doubt, chronic impostor feelings contribute to anxiety, depression and burnout \cite{bernard2002applying, sakulku2011impostor, tulili2023burnout, al2025prevalence}. 
% Within SE development environments, where performance is continually assessed through code reviews, issue tracking, and peer feedback, these dynamics may considerably intensify. 
When psychological safety \cite{edmondson1999psychological} is low, individuals often suppress questions or contributions, reinforcing the impostor feelings. Consequently, IP should be understood not merely as an individual trait but as a cooperative and contextual phenomenon embedded in the software engineering environment.

%Besides, a previous study \cite{Guenes24} involving software engineers revealed that more than half of software engineers experience frequent to intense IP levels, negatively impacting their perceived productivity. Moreover, underrepresented groups suffer disproportionately \cite{oliveira2024navigating}, leading to relevant negative impacts on their well-being \cite{Guenes25}. 

%From a theoretical standpoint, IP can be interpreted as a deficit in self-efficacy \cite{bandura2013self} and an outcome of maladaptive social comparison \cite{arigo2024social}. Within collaborative development environments, where performance is continually assessed through code reviews, issue tracking, and peer feedback, these dynamics may considerably intensify. When psychological safety is low \cite{edmondson1999psychological}, individuals often suppress questions or contributions, reinforcing impostor cycles. Consequently, IP should be understood not merely as an individual trait but as a cooperative and contextual phenomenon embedded in the social fabric of SE work.

Given the high prevalence and negative consequences of IP, there is a pressing need to identify practical and evidence-based mitigation strategies \cite{para2024interventions}. Among the approaches discussed in the broader psychological literature, professional coaching appears particularly relevant because it targets both the cognitive and affective mechanisms that sustain IP \cite{magro2022hiding}. Indeed, evidence from other professional domains already suggests that coaching can reduce fear of negative evaluation and encourage more adaptive self-attributions \cite{zanchetta2020overcoming, evers2006quasi}. However, despite its potential in other fields, no prior research has experimentally evaluated coaching interventions for IP in SE. 

Addressing this gap, our study empirically investigates whether a structured group-coaching program can mitigate IP feelings among early-career software engineers. This population is particularly vulnerable; impostor experiences are often heightened during the transition from education to professional practice \cite{sakulku2011impostor}, creating a considerable internal barrier to career development \cite{neureiter2016inner}. This is especially relevant in SE, where prior studies show that novice professionals report higher impostor scores and lower perceived productivity than their more experienced peers \cite{maji2021they, oliveira2024navigating}.

% Responding to this gap, our study aims to empirically examine whether a structured group-coaching program can mitigate impostor feelings among early-career software engineers. We focused on this population because impostor experiences are particularly pronounced during the transition from education to professional practice \cite{sakulku2011impostor} and the impact of the IP as a considerable inner barrier to career development. Moreover, prior studies have shown that novice software professionals report higher impostor scores and lower perceived productivity than their more experienced peers \cite{maji2021they, Guenes24, oliveira2024navigating}.

To investigate this phenomenon under structured yet realistic conditions, we conducted a quasi-experiment with 20 early-career software engineers engaged in a technological innovation program that involved real-world clients and external funding. This approach enabled us to study the effects of a psychological intervention in a team-based environment while maintaining methodological consistency. Random assignment of individuals was not feasible (participants belonged to pre-existing project teams), therefore we applied a wait-list control design, ensuring that all participants received the coaching intervention while preserving comparability. The study took place in a university-affiliated innovation lab.

Impostor feelings are known to co-occur with broader psychological states (such as reduced well-being \cite{Guenes25}, emotional strain \cite{al2025prevalence}, and lower life satisfaction \cite{islam2024effect}), hence we complemented the measurement of IP with evaluated instruments that capture these related dimensions. This approach allowed us to explore whether changes in impostor scores were accompanied by changes in participants’ affective experience and self-perceived well-being, offering a comprehensive overview of the intervention’s a effects and effectiveness. To assess these effects, we used the Clance Impostor Phenomenon Scale (CIPS) alongside instruments for well-being (WHO-5), life satisfaction (SWLS), and affect (PANAS). To complement the quantitative measures, we also employed non-participant observation \cite{ostrower1998nonparticipant} throughout the sessions, both to ensure fidelity to the coaching protocol and to obtain contextual nuances in group interactions and engagement.

%To assess these effects, we measured impostor feelings using the Clance Impostor Phenomenon Scale (CIPS) and complemented them with validated instruments for well-being (WHO-5), life satisfaction (SWLS), and affect (PANAS). Quantitative data were triangulated with field observations documenting group dynamics and engagement during the sessions.

%Despite this promise, the broader literature on IP interventions suffers from a lack of methodological rigor. A recent systematic review highlighted this gap, calling for future research to move beyond anecdotal advice and adopt robust research designs, such as quasi-experimental studies with systematic pre- and post-intervention measurements \cite{para2024interventions}. In response to this call, our study evaluates a group-coaching intervention within a software engineering context, employing a quasi-experimental design with pre- and post-intervention measures to assess its impact.

Overall, our findings indicate that while the treatment group experienced a reduction in impostor feelings, the control group also exhibited a decrease during the observation period. This observation suggests that contextual and organizational factors (e.g. as team climate, peer validation, or project stage) may play a more dominant role than the formal intervention in reducing these feelings.

%Furthermore, we incorporated non-participant observation \cite{ostrower1998nonparticipant}, a foundational qualitative research method, for two key purposes. First, we used it to ensure intervention fidelity by verifying that the same coaching protocol was delivered consistently to all groups. Second, it served as our primary method for gathering rich qualitative data on the intervention process, allowing us to observe participant engagement, assess group dynamics, and identify contextual factors that could influence the study's outcomes.

% However, despite its theoretical alignment, the efficacy of structured coaching as a targeted intervention for IP has not, to our knowledge, been empirically evaluated within the software engineering domain.

% coaching to self-efficacy \cite{evers2006quasi}

In summary, this paper offers the first empirical examinations of a coaching-based intervention aimed at mitigating impostor feelings in SE. This study also provides an account of how the intervention unfolded in a team environment and what was learned from its application. Beyond reporting effects, we reflect on contextual factors (for example, team dynamics and feedback culture) that potentially influenced the outcomes.

\section{Background and Related Work}
\label{sec:background}
%This section reviews the conceptual and empirical foundations of IP, its specific relevance to SE, and the rationale for exploring coaching as a psychologically grounded intervention.

%This section provides the background on the Impostor Phenomenon, IP as a challenge for software engineers and the Coaching intervention as a possible solution.

\subsection{The Impostor Phenomenon in Software Engineering}

The Impostor Phenomenon, first identified by Clance and Imes \cite{clance1978imposter}, describes a pervasive internal experience of intellectual fraudulence despite clear evidence of competence. Individuals who experience IP find it difficult to accept success as legitimate, often attributing accomplishments to luck, timing, or a belief that others are overestimating that person's ability. Clance later conceptualized this pattern as an Impostor Cycle \cite{clance1985impostor}, wherein anxiety before a performance task leads to over-preparation or procrastination; subsequent success is discounted, reinforcing self-doubt and preventing the consolidation of a stable sense of competence. Over time, this loop undermines self-confidence, fuels perfectionism, and sustains a chronic fear of exposure.

Extensive research has linked IP to a range of negative outcomes, including anxiety, depression, burnout, and decreased well-being \cite{villwock2016impostor, fimiani2021interpersonal, cawcutt2021bias}. These effects are often mediated by low self-efficacy, maladaptive attributional styles, and heightened fear of negative evaluation \cite{gullifor2024impostor}, all of which distort how individuals interpret feedback and success. Beyond individual distress, IP impairs learning and collaboration by discouraging open communication and help-seeking, thereby constraining psychological safety within teams.

In the context SE, IP is characterized by distinctive features. The constant demand for expertise in rapidly evolving technologies makes SE a setting prone for comparison, often amplifying self-doubt and impostor feelings~\cite{maji2021they, oliveira2024navigating}. Large-scale empirical studies already revealed that among more than 600 software professionals across 26 countries, over half reported frequent or intense impostor feelings \cite{Guenes24}. Gender and racial disparities are pronounced, with higher prevalence among women and especially among Black women in technical roles \cite{Guenes25}. These findings mirror broader organizational inequities, linking IP to inclusion and belonging. Moreover, IP tend to decline with experience but often resurface at transitional moments suggesting that impostorism is not a passing phase but a recurrent psychological challenge \cite{sakulku2011impostor}.

To quantify IP, the Clance Impostor Phenomenon Scale (CIPS) \cite{CIPSclance1985impostor} remains the field’s standard instrument\cite{mak2019impostor}, capturing the frequency of impostor thoughts and behaviors. However, IP rarely acts in isolation. It intertwines with emotional well-being, life satisfaction, and affective balance \cite{mcelwee2007feeling}, dimensions essential to understanding how self-doubt manifests and evolves. Three evaluated scales are frequently used to examine these dimensions. The WHO-5 Well-Being Index \cite{nylen2022validating} measures subjective vitality and positive mood, offering a concise snapshot of psychological well-being over recent weeks; low scores often accompany stress or depressive states. The Satisfaction With Life Scale (SWLS) \cite{diener1985satisfactionswls} assesses the cognitive dimension of well-being, including how individuals evaluate their overall life circumstances relative to their aspirations. Finally, the Positive and Negative Affect Schedule (PANAS) \cite{watson1988developmentpanas} distinguishes between positive affect (enthusiasm, engagement) and negative affect (distress, guilt, irritability), providing a fine-grained view of the affective tone. These instruments situate IP within the broader architecture of psychological functioning: CIPS captures cognitive self-doubt, WHO-5 reflects vitality, SWLS reveals life appraisal, and PANAS traces emotional balance. 

Beyond cognition and affect, IP is deeply entangled with identity and belonging. In team-based professions like SE, success is not only individual but socially constructed through peer recognition and collective norms \cite{teh2011social, hoda2021socio}. When individuals perceive a mismatch between their self-concept and the community’s image of competence, IP can intensify. Studies of underrepresented professionals show that a lack of role models, stereotype threat, and limited access to affirming feedback potentialize impostor experiences \cite{oliveira2024navigating}.

\subsection{Coaching as an Intervention}
Coaching is a structured and dialogical process in which a trained professional facilitates reflection, learning, and personal growth through purposeful conversation \cite{grant2003impact}. Unlike mentoring or therapy, coaching does not rely on the transfer of expertise or clinical treatment; instead, it encourages self-awareness and self-regulation \cite{palmer2007practice}. Drawing on cognitive-behavioral and humanistic traditions, coaching helps individuals confront self-limiting beliefs and strengthen confidence in their own abilities \cite{allen2016roots}.

The principles that underlie coaching correspond closely to those implicated in the IP \cite{zanchetta2020overcoming}. Through cognitive reframing, participants are guided to reinterpret success and failure in ways that reinforce internal attributions of competence \cite{norbury2025learning, macgabhann2025some}. Coaching also promotes emotional regulation, equipping individuals to manage anxiety, perfectionistic tendencies, and fear of negative evaluation \cite{richter2021positive, junker2021impact}. When conducted in groups, it introduces a social dimension that allows participants to share experiences, identify recurring patterns of self-doubt, and collectively reframe their perceptions \cite{pannell2021adult, mbokota2022role}. These exchanges can normalize vulnerability and strengthen psychological safety, counteracting the isolation that often amplifies impostor feelings \cite{para2024interventions}.

Empirical studies consistently indicate that coaching contributes to improved goal attainment, self-efficacy, and well-being \cite{theeboom2014does, de2023can}. In this regard, interventions emphasizing structured reflection and feedback have been found to enhance internal locus of control and emotional stability, both inversely related to impostor tendencies \cite{rotter1954social, evers2006quasi}. Within IP research, Zanchetta \textit{et al.} \cite{zanchetta2020overcoming} showed that a coaching program for young professionals significantly reduced impostor scores, primarily through a reduction in fear of negative evaluation and an increase in self-enhancing attributions. A subsequent review of 31 intervention studies reached similar conclusions, identifying group-based programs that combine cognitive reframing with supportive dialogue as among the most effective approaches for mitigating impostor experiences \cite{para2024interventions}.

These converging findings suggest that coaching may be potentially suited to SE contexts, where performance visibility and rapid change on are intrinsic to professional practice. However, despite theoretical alignment and encouraging evidence from other fields, no experimental studies have yet examined whether coaching can mitigate impostor feelings in SE.

% explicar oq é coaching

%The selection of coaching as a potential intervention to mitigate IP is grounded in substantial evidence of its efficacy in professional settings. A recent scoping review of 31 IP intervention studies identified group-based support as a primary and effective lever for change, which motivates our choice of a group format \cite{para2024interventions}. More specifically, a comprehensive meta-analysis of 37 Randomized Controlled Trials (RCTs) demonstrated that workplace coaching has a statistically significant and moderately strong positive effect (g = .59) on a range of personal outcomes, particularly when conducted by qualified coaches \cite{de2023can}.

%Most relevant to this study, a randomized controlled trial by Zanchetta et al.  \cite{zanchetta2020overcoming} specifically tested a coaching intervention for IP with 103 young employees. Their results revealed that coaching was an effective intervention for sustainably reducing IP scores. The study identified that this effect was significantly mediated by a reduced "fear of negative evaluation" and that the intervention also improved participants' self-efficacy and self-enhancing attributions. As the specific intervention materials used by Zanchetta et al. are not publicly available, a direct replication was not feasible. Therefore, we designed a novel group coaching intervention tailored specifically to address this problem within the context of early-career software engineering professionals.
\section{Experimental Design}
\label{sec:experimentaldesign}
% incluir as outras escalas e o non participant obervevable

To examine whether structured coaching can mitigate IP among early-career software engineers, we conducted a quasi-experiment following the recommendations by Wohlin \textit{et al.}  \cite{wohlin2024experimentation}. The intervention under scope, developed by a certified coach in collaboration with psychology researchers, aimed to assess the effects of group coaching on impostor feelings and related psychological outcomes. Participants completed validated instruments measuring the IP (CIPS), subjective well-being (WHO-5), life satisfaction (SWLS), and affective balance (PANAS) before, during, and after the intervention to monitor potential changes over time. In addition to these quantitative measures, the study incorporated non-participant observation \cite{ostrower1998nonparticipant} to document group dynamics, engagement, and adherence to the coaching protocol. Below, we clarify the main phases of our experiment process.

%Figure~\ref{fig:experiment-planning-figure} summarizes the main phases of the experimental planning process.

%An independent observer attended all sessions, taking detailed field notes to ensure consistency in delivery and to capture contextual factors that could influence participants’ responses. This qualitative perspective provided a complementary layer of evidence.

%A quasi-experiment was conducted with students enrolled in a technological innovation training program. An intervention inspired by research from the area of Psychology was designed by a certified coach to evaluate whether coaching could reduce the impostor feelings of software professionals. Levels of IP were assessed using a validated scale before, during, and after the coaching sessions to monitor changes over time. The experimental study plan steps are detailed the experimental as suggested by Wohlin et al. [24] (Figure \ref{fig:experiment-planning-figure}),

%\begin{figure}[ht!]
%\includegraphics[width=0.9\linewidth]{Images/experiment-planning-figure.png}
%  \caption{Experimental planning steps}
 % \Description{Planning phase overview}
  %\label{fig:experiment-planning-figure}
%\end{figure}

% Controlled experiment
% Coaching interventions aimed at strengthening

% self-confidence
% self-efficacy
% self-esteem

\subsection{Goal Definition}
Following the Goal–Question–Metric (GQM) goal definition template \cite{caldiera1994goal}, the study objective was defined as follows: \textit{Analyze} a structured group-coaching intervention \textit{for the purpose of} characterization \textit{with respect to its} impact in reducing the Impostor Phenomenon (CIPS) levels and influencing related psychological indicators, well-being, life satisfaction, and affect (WHO-5, SWLS, PANAS), \textit{from the point of view} of the researcher, \textit{in the context} of a quasi-experiment involving early-career software engineers from two pre-existing project teams of an industry-academia collaboration within an university-affiliated lab.

%\textit{Analyze} a structured group-coaching intervention \textit{for the purpose of} evaluation \textit{with respect to} its effectiveness in reducing Impostor Phenomenon (IP) levels and its influence on related psychological indicators of well-being, life satisfaction, and affect (WHO-5, SWLS, and PANAS) \textit{from the point of view of} the researcher in the context of a quasi-experiment with early-career software engineers working in two pre-existing project teams within a university-affiliated lab.

%The goal was defined using GQM as follows:
%Analyze a group coaching intervention
%for the purpose of characterization
%with respect to its effectiveness in reducing the levels of Impostor Phenomenon (IP)
%from the point of view of the researcher
%in the context of a quasi-experiment with early-career software engineers.

% sublinhar as palavras do GQM (analyze)...

\subsection{Context Selection}
As outlined below, the experimental context was defined according to four classical design dimensions proposed by Wohlin \textit{et al.} \cite{wohlin2024experimentation}:
%As clarified below, the experimental context was defined along four classical dimensions of design \cite{wohlin2012experimentation} (delivery format, participant profile, problem realism, and contextual specificity) to ensure transparency regarding the study’s scope and implications for validity.

%Following the goal definition, the context for this research was characterized along four key dimensions—off-line vs. online, students vs. professionals, real vs. toy problem, and specific vs. general—to ensure clarity regarding the experiment's environment and its implications for validity.
\begin{itemize}
    \item \textit{Off-line vs. On-line}: The intervention was delivered through in-person group coaching sessions. To minimize potential reactivity and demand characteristics, the activity was presented to participants as part of their existing training program rather than as a standalone session. This integration allowed the sessions to unfold within a natural and familiar learning environment while preserving rigor over the session structure and delivery fidelity.
    
    %The intervention was delivered through in-person coaching sessions. To mitigate participant reactivity and demand characteristics, the intervention was intentionally framed as part of the participants' existing program. It was presented not as a separate experiment, but as one of the diverse training modules they were scheduled to receive. This in-person approach was chosen to ensure a controlled environment and maintain intervention fidelity (i.e., consistent delivery of the protocol).
    \item \textit{Students vs. Professionals}: 
    Participants were late-stage undergraduate students enrolled in a technological innovation program. We selected this context specifically to capture the critical transition from education to professional practice. Prior research identifies this as a period when individuals are highly vulnerable to IP, as they face new performance expectations and identity shifts \cite{sakulku2011impostor}. Our participants were not typical students; their role was that of early-career software engineers, requiring them to manage collaborative projects and engage with real-world industry clients. This provided an environment with a suitable setting to examine IP. Therefore, while findings may not generalize to experienced practitioners, they offer a precise window into how IP manifests and is mitigated at the onset of a software engineering career.
    % Participants were undergraduate students enrolled in a technological innovation training program, serving as credible proxies for early-career software engineers. This choice, while limiting direct generalization to industry professionals, provided a ecologically relevant setting to examine a real psychological phenomenon under conditions of collaborative project work. Prior research shows that IP is pronounced during the transition from education to professional practice, when individuals face new performance expectations and identity shifts \cite{sakulku2011impostor}. 
    %The participants’ ongoing engagement in applied software projects and frequent interaction with external partners mirrored the responsibilities and social dynamics of entry-level professionals, making this cohort especially suitable for our study. 
    
    %The subjects of this experiment were students enrolled in a technological innovation training program. While the ultimate goal is to address IP in software professionals, using early-career students as subjects offered a practical and controlled setting. This choice represents a trade-off, prioritizing internal validity and feasibility over the immediate generalizability to industry professionals. The participants, being in a technology-focused program, serve as a suitable proxy for early-career professionals.
    \item \textit{Real vs. Toy Problem}: The participants worked on industry–academia collaboration projects (with real-world clients and funding) that required delivering concrete technological solutions to external partners, reproducing the pressures and expectations typical of professional practice. The intervention itself was a structured group-coaching program, co-designed with psychology researchers and delivered by a certified coach.% following the same principles and structure used in workplace settings. %This configuration allowed situating the investigation of impostor feelings within a context that closely resembled early professional experience.
    
    %The experiment addresses a real problem. Although situated in an academic environment, the focus of the study, which is the psychological impact of IP and its mitigation, is a genuine and significant issue faced by professionals in the software industry. The intervention itself, a coaching program designed with Psychology researchers and delivered by a certified coach, is a real-world strategy, not a simplified or simulated task.
    \item \textit{Specific vs. General}: 
    The study was conducted within a single cohort of participants enrolled in a defined institutional program, resulting in a specific research context that allowed for consistency in conditions and procedures, though without full experimental control. While this scope limits the generalizability of the findings, it strengthens internal coherence and facilitates transparent analysis. %Such designs are common in quasi-experimental research, where establishing clear and context-specific evidence is a necessary step toward future replications in more varied professional environments.
    
    %The study was carried out within a single cohort of participants enrolled in a defined institutional program, resulting in a specific and well-controlled research context. While this bounded scope constrains the generalizability of the findings, it enhances internal validity and ensures consistency across participants and procedures. Such focused designs are characteristic of quasi-experimental studies that seek to produce initial, reliable evidence under clearly delineated conditions, serving as a foundation for subsequent replications in broader and more diverse professional settings.
    
    %The context is specific. The findings are derived from a single cohort of early-career software engineers within one particular training program. This specificity allows for high control over the experimental conditions.
    
    % While the results will be highly relevant to this specific context, further research will be required to generalize the findings to the broader population of software professionals across different environments and experience levels.
\end{itemize}

The sample consisted of 20 participants, organized into two project teams (P1 and P2) of ten members each. These teams were subdivided into two stable sub-groups of five (G1 and G2 for P1; G3 and G4 for P2) to facilitate group dynamics and ensure manageable coaching interactions.
Both projects were developed in partnership with external organizations and required the delivery of functioning features to real clients, ensuring practical relevance. The teams operated under the same organizational structure and management framework, following agile principles and the full cycle of Scrum ceremonies. 

\begin{itemize}
    \item \textit{Project 1 – Automation (P1)}: relates to the automation of energy transmission diagram transcription and was technically more complex, requiring the development of an interpreter to convert DXF files into ASCII format. The system replaced a manual, labor-intensive process for the customer, demanding higher levels of algorithmic design and data manipulation.

\item \textit{Project 2 – Predictive Analytics (P2)}: focused on building a predictive model for accident prevention in oil and gas operations. Using historical data, the team implemented a prototype to estimate risk probabilities, incorporating contextual factors such as weather and workload. Compared to P1, this task involved less technical complexity but equivalent collaboration and delivery expectations.
\end{itemize}

% \begin{itemize}

% \item \textit{Project 1 – Automation (P1)}. This project involved the automation of energy transmission diagram transcription and was technically more complex, requiring the development of an interpreter to convert DXF files into ASCII format. The system replaced a manual, labor-intensive process for the client, demanding higher levels of algorithmic design and data manipulation.

% \item \textit{Project 2 – Predictive Analytics (P2)}.
% This project focused on building a predictive model for accident prevention in oil and gas operations. Using historical data, the team implemented a prototype to estimate risk probabilities, incorporating contextual factors such as weather and workload. Compared to P1, this task involved less technical complexity but equivalent collaboration and delivery expectations.
% \end{itemize}

Both teams had comparable compositions in size, supervision, disciplinary background, and experience level, and they worked concurrently under the same lab conditions and deadlines. The asymmetry in project complexity was acknowledged during study design and monitored through observation, but both projects provided industry-connected tasks aligned with early professional practice. This parallel organization ensured a consistent context for implementing the quasi-experimental wait-list control design.

\subsection{Hypothesis Formulation}

%We formulated the following directional hypotheses:

Given the exploratory nature of this quasi-experiment and the limited sample size, our hypothesis was formulated to ground directional expectations rather than formal statistical tests:
\begin{itemize}
    \item (H1) Participation in the group-coaching intervention will be associated with a reduction in the Impostor Phenomenon (CIPS) scores.
    
    %(H1) Participation in the group-coaching intervention will be associated with a decrease in Impostor Phenomenon (CIPS).
    \item (H2) The intervention will be associated with improved psychological well-being (WHO-5) and life satisfaction (SWLS), an increase in positive affect, and a decrease in negative affect (both measured by PANAS).
\end{itemize}

% The expectation was that \textbf{participation in the group-coaching intervention would be associated with a decrease in IP} and \textbf{potential improvements in indicators of well-being, life satisfaction, and affect when compared to the observation period without coaching}. 

The analysis focused on examining both the direction of score changes (whether participants’ scores moved in the expected direction after the intervention) and their magnitude, that is, the extent of differences across time points. Following recommendations for small-sample and exploratory designs \cite{wohlin2024experimentation, cumming2014new, lakens2013calculating}, we prioritized descriptive and effect-based interpretation over formal significance testing. This approach emphasizes estimation and trend identification rather than dichotomous inference, enabling a more nuanced understanding of potential intervention effects. The primary construct under investigation was the Impostor Phenomenon, measured using the CIPS, complemented by contextual insights from the WHO-5, SWLS, and PANAS scales.

%The analysis focused on examining both the direction of these changes (whether participants’ scores moved in the expected direction after the intervention) and their magnitude, that is, the extent or size of the observed differences across time points. This approach allowed us to assess whether the coaching produced meaningful trends consistent with theoretical expectations, even without relying on large-sample significance testing. The primary construct under investigation was the IP, measured using the CIPS, complemented by contextual findings from WHO-5, SWLS, and PANAS results.

%To formally evaluate the effectiveness of the coaching intervention, we formulated a null hypothesis ($H_0$) and an alternative hypothesis ($H_1$). The null hypothesis posits that the intervention has no statistically significant effect on reducing the levels of the impostor phenomenon among participants. Conversely, the alternative hypothesis states that the coaching is effective, leading to a measurable reduction in IP levels. Formally, let $\mu_{\text{coaching}}$ be the mean IP level for the treatment group and $\mu_{\text{control}}$ be the mean IP level for the control group. The hypotheses are stated as:

%\begin{itemize}
 %   \item \textbf{Null Hypothesis ($H_0$):} $\mu_{\text{coaching}} = \mu_{\text{control}}$
  %  \item \textbf{Alternative Hypothesis ($H_1$):} $\mu_{\text{coaching}} < \mu_{\text{control}}$
%\end{itemize}

\subsection{Variables Selection}
The independent variable in this quasi-experiment was the group-coaching intervention, a categorical factor with two levels: participation in the coaching sessions (treatment condition) and non-participation during the observation period (control condition). The primary dependent variable was the intensity of IP, measured using the CIPS, which produces a continuous score representing the frequency and strength of impostor feelings. Complementary dependent measures captured related psychological dimensions: subjective well-being (WHO-5), life satisfaction (SWLS), and affective balance (PANAS). These scales allowed for a multidimensional view of potential changes in participants’ psychological states, enabling descriptive comparison of trends across time points rather than reliance on inferential statistics.

%The primary independent variable in this study is the coaching intervention. This is a categorical variable with two levels that define the experimental groups: the treatment group, which participated in the coaching sessions, and the control group, which did not receive the intervention during the observation period. The dependent variable, derived directly from our hypothesis, is the level of IP. This construct was measured indirectly using an evaluated scale, resulting in a numerical score that represents the intensity of each participant's impostor feelings.

\subsection{Selection of Subjects}
Participants worked within pre-existing project teams; therefore, individual random assignment was not feasible without disrupting established workflows and team cohesion. Instead, we adopted a non-probability convenience sampling approach, selecting two comparable teams of early-career software engineers from the same university-affiliated innovation lab. 
% repeted in 3.2 Ensuring comparable exposure to contextual factors, both teams operated under similar supervision, project timelines, and organizational structures, though they differed slightly in technical complexity. 

% The study subjects (N=20) comprised late-stage undergraduate students from a university laboratory renowned for its strong industry ties. Participants' ages ranged from 19 to 26, with most (N=15) enrolled in Computer Engineering or Computer Science. While for many this lab represented their first significant professional engagement, two participants (10\%) already possessed over three years of external software engineering experience. Regarding ethnic diversity, the group was composed of 19 white participants and one black participant. This laboratory simulates a corporate environment where participants function as an applied development team, holding regular meetings with external industry customers and delivering a final product. The fidelity of this environment as an industry pipeline is underscored by the fact that some participants were invited for positions at the partner companies immediately following the project's conclusion. 

Subjects ($N=20$) were late-stage undergraduate students (ages 19–26) from a university laboratory with strong industry ties. Most ($N=15$) were enrolled in Computer Engineering or Science. While this was the first professional engagement for most, two participants (10\%) had over three years of external software engineering experience. The cohort was ethnically composed of 19 white and one black participant. The laboratory simulates a corporate environment where participants function as an applied development team for external customers; its high fidelity is evidenced by participants receiving job offers from partner companies immediately upon project conclusion.

%Project 1 (P1) was designated as the initial treatment group, while Project 2 (P2) served as the wait-list control. This sequencing allowed for ethical balance (ensuring that all participants received the coaching) while supporting temporal comparison between coached and uncoached phases. Each team’s participation in industry-linked projects provided a realistic work environment that closely mirrored early professional practice.

%In applied software engineering research, it is often infeasible to randomly assign individuals from pre-existing teams, as this would disrupt project integrity and team dynamics. Therefore, for this quasi-experiment, we used a non-probability sampling technique, specifically convenience sampling. The subjects were early-career software engineers from two project teams within the same lab.

%Therefore, the two conveniently sampled teams were non-randomly assigned to their roles. Project 1 served as the initial treatment group, and Project 2 served as the wait-list control group. To ethically ensure fairness, the control group received the same coaching intervention after the treatment group's final measurements were complete.

\subsection{Choice of Design Type}
We followed a one-factor and two-treatment quasi-experimental design \cite{wohlin2024experimentation}, incorporating a wait-list control. The single factor under investigation was participation in the coaching intervention, with two treatment conditions: receiving coaching (experimental treatment) and not receiving coaching during the observation period (control). P1 served as the initial treatment group, receiving the three-session coaching intervention first, while P2 functioned as the wait-list control group, completing the same sequence of measurements before later receiving the identical intervention. Blocking was applied at the project level to account for potential inter-team variability, such as task complexity or team culture, while the unit of analysis remained the individual participant. Although full randomization was not feasible, the order of intervention between teams was randomized to mitigate sequencing bias, and group sizes were equivalent. 
% repeated in 3.5 Both teams operated concurrently under the same supervision and schedule, ensuring comparable exposure to contextual factors. 
The same certified coach delivered all sessions using an identical protocol, and a trained observer (the first author) attended every session to verify intervention fidelity and record contextual observations. Data were collected at four time points (T0–T3), allowing examination of within-group and between-group trends across the intervention phases.

%Our experimental design applied the principles of randomization, blocking, and balancing. To control for inter-project variability, participants were blocked by project. 
% Grouping teams by project allowed us to isolate the effects of confounding variables specific to each project, like code complexity and team culture, from the final results.
%The treatment order was randomized, and the design was balanced with an equal number of participants in each group. 
%The goal was to investigate weather coaching interventions reduces the impostor feelings. 
%The dependent variable is the level of Impostor Phenomenon. To achieve this research goal, we employed a one-factor, two-treatment experimental design \cite{wohlin2024experimentation}. The experiment was structured around a common subject (a series of documented coaching sessions) which was subjected to two distinct treatments: receiving coaching (the treatment group) and not receiving coaching (the control group).

\subsection{Instrumentation}
This study combined a questionnaire using validated scales to measure constructs of interest, structured observation, and a theoretically grounded coaching protocol to capture both quantitative changes and qualitative dynamics across the four time points (T0–T3). The instrumentation was co-designed by the SE and Psychology research teams. The questionnaire was applied online via a Tally platform, ensuring consistent administration and data protection. Before participation, all subjects provided informed consent through an embedded form outlining the research goals, procedures, confidentiality assurances, and their right to withdraw at any time. The protocol received formal approval from the university’s ethics board and adhered to ethical standards commonly adopted in empirical SE studies \cite{singer2003ethical}. After consenting, participants completed a short demographic questionnaire to record age, academic background, and prior professional experience, enabling contextual interpretation of individual differences.

The psychometric core of the instrument comprised four evaluated scales capturing complementary psychological constructs. The Clance Impostor Phenomenon Scale (CIPS) served as the primary measure of IP. It consists of 20 statements describing impostor-related thoughts and behaviors (e.g., “I’m afraid people important to me may find out that I’m not as capable as they think I am”), rated on a five-point Likert scale ranging from 1 (“Not at all true”) to 5 (“Very true”). Total scores range from 20 to 100, where higher values indicate stronger impostor experiences. Following Clance’s interpretive guidelines, scores $\leq$ 40 represent few impostor feelings, 41–60 moderate, 61–80 frequent, and > 80 intense \cite{clance1985impostor, mak2019impostor}. CIPS was the main dependent variable for assessing direction and magnitude of change across measurement points.

To situate IP within broader psychological functioning, we included three additional scales capturing emotional well-being, cognitive satisfaction, and affective balance. The World Health Organization Five Well-Being Index (WHO-5) \cite{nylen2022validating} assesses positive mood and vitality over previous two weeks using five short statements (e.g., “I have felt cheerful and in good spirits”). Each item is rated from 0 (“At no time”) to 5 (“All of the time”), producing a raw score of 0–25 that is multiplied by four to yield a standardized score from 0 to 100. Higher values indicate greater well-being, while scores below 50 suggest low psychological well-being and possible depressive symptoms. The WHO-5 is sensitive to short-term changes and has been widely validated as an index of emotional vitality.

The Satisfaction With Life Scale (SWLS) \cite{diener1985satisfactionswls} was used to measure the cognitive dimension of subjective well-being. It includes five evaluative statements such as “In most ways my life is close to my ideal”, rated on a seven-point scale from 1 (“Strongly disagree”) to 7 (“Strongly agree”). Summed scores range from 5 to 35, with higher values representing greater satisfaction with life. Standard interpretive bands classify scores from 20–24 as neutral, 25–29 as slightly satisfied, and 30–35 as highly satisfied. The SWLS captures reflective life evaluations rather than transient affective states, thus complementing the other instruments.

The Positive and Negative Affect Schedule (PANAS) \cite{watson1988developmentpanas} assessed affective balance across two dimensions: positive affect (PA) and negative affect (NA). Participants rated twenty emotion words, ten positive (e.g., “Enthusiastic”, “Inspired”) and ten negative (e.g., “Distressed”, “Upset”), indicating to what extent they experienced each during the past week, using a five-point scale from 1 (“Very slightly or not at all”) to 5 (“Extremely”). The two subscales (PA and NA) are scored separately, each ranging from 10 to 50. Rather than computing a difference score, we analyzed PA and NA trajectories independently. PANAS captures short-term affective shifts that complement the longer-term indicators of WHO-5 and SWLS. %All four scales were administered at each data collection point (T0–T3).%, maintaining identical ordering and wording to ensure comparability. 

In addition to the quantitative instruments, the study incorporated non-participant observation to ensure intervention fidelity and to capture contextual nuances of group behavior during the coaching sessions. An observer attended each session without interacting with participants, following a structured checklist to verify adherence to the planned coaching content, time allocations, and facilitation sequence. Complementary field notes documented participant engagement, interpersonal dynamics, and moments of reflection. These observations were not analyzed quantitatively but served two methodological purposes: confirming that both project teams received the intervention consistently and enriching the interpretation of quantitative trends, particularly when changes emerged in the absence of formal treatment.

A last and essential component of the instrumentation was the intervention protocol itself, which served as a standardized treatment instrument. The protocol was designed by the Psychology research team, including a certified coach, drawing upon a recent evidence-based synthesis of the IP psychological antecedents \cite{gullifor2024impostor}. Its structure was grounded in cognitive-behavioral and humanistic theories, focusing on five constructs known to support impostor feelings: self-esteem, self-efficacy, internal locus of control, emotional stability, and attributional style. The coaching was planned into a structured series of three two-hour group coaching sessions, to be conducted weekly for a total exposure of six hours per participant. Each session emphasized one or more of the five theoretical constructs supporting the intervention and operationalized through reflective dialogue, guided exercises, and peer feedback. Session 1 focused on recognizing impostor patterns and strengthening self-awareness; Session 2 emphasized cognitive reframing and the development of self-efficacy and internal control; and Session 3 addressed emotional regulation and adaptive attributional styles, encouraging participants to integrate insights into their professional practice. Between sessions, participants completed brief reflection tasks designed to reinforce learning and promote behavioral experimentation.

%hiding the paragraph below to gain some space
%Self-esteem refers to one’s overall sense of self-worth and competence \cite{ziller1969self}. Individuals with low self-esteem tend to interpret success as accidental and failures as personal evidence of inadequacym an attributional pattern key to impostorism. Self-efficacy, derived from Bandura’s social-cognitive theory \cite{bandura1986social}, denotes the belief in one’s capacity to perform tasks and overcome challenges. High self-efficacy buffers against impostor thoughts by reinforcing internal evidence of competence. Internal locus of control, represents the perception that outcomes are primarily the result of one’s own actions rather than external circumstances \cite{rotter1954social}. Those with a strong internal locus of control are less likely to attribute achievements to luck or external factors, thereby reducing self-doubt. Emotional stability involves the ability to regulate emotions and tolerate stress without excessive anxiety or self-criticism \cite{li2016emotional}. Because impostor experiences are often intensified by perfectionism and fear of failure, fostering emotional stability helps participants manage evaluation anxiety and maintain perspective. Finally, attributional style, as conceptualized by Weiner \cite{weiner1985attribution}, describes that more adaptive attributional style encourages individuals to view success as internally caused and failure as a temporary, controllable event, directly countering the maladaptive attributions typical of impostor thinking.
\vspace{-3pt}
\subsection{Data Collection and Operation}

Prior to data collection, participants were informed that they would take part in a structured professional development program as part of their innovation training curriculum, without explicit mention IP to avoid priming effects. Participants knew all data would be analyzed anonymously. The coach and research team jointly conducted a procedural alignment meeting to standardize delivery and clarify timing, instrumentation, and observer responsibilities. 
% No pilot or dry-run session was required, as the coaching protocol had been previously validated in a small internal workshop and the instruments were standardized.

\begin{figure}[htbp]
\centerline{\includegraphics[width=\linewidth]{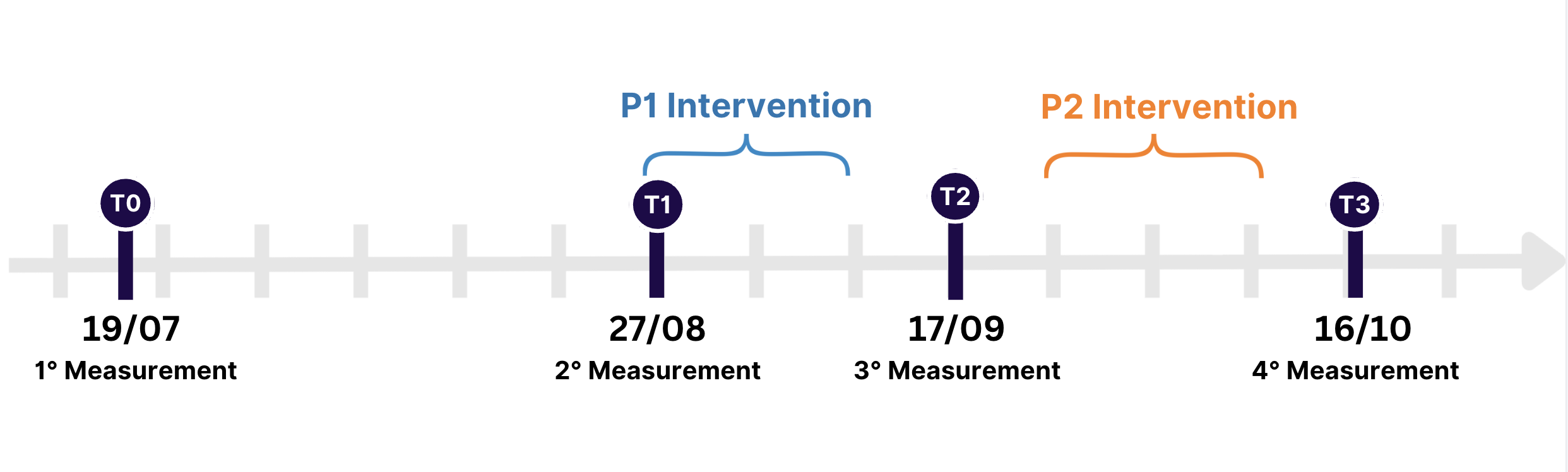}}
    \caption{Timeline of measurements and intervention phases. T0 and T1 establish the pre-intervention baseline; T2 and T3 represent post-intervention measures for P1 and P2, respectively.}
    \Description{Timeline}\vspace{-10pt}
    \label{fig:intervention_timeline}
\end{figure}

As can be seen in Figure \ref{fig:intervention_timeline}, at T0 (baseline), around five weeks before the first intervention phase, all participants (P1 and P2) completed the full battery of instruments to establish baseline scores. This pre-intervention period (T0 to T1) was intended to ensure responses were stable before the intervention began. At T1, immediately before the first coaching session, participants from both projects again completed the same measures to assess any natural fluctuations prior to treatment. The first intervention phase then occurred over three consecutive weeks, during which the treatment group (P1) received the coaching sessions while the control group (P2) continued its regular project work. At T2 (post-intervention for P1/pre-intervention for P2), both teams completed the questionnaires simultaneously: P1’s data served as its post-treatment measure, while P2’s responses constituted its pre-treatment baseline. Finally, at T3 (post-intervention for P2), the second intervention phase concluded with the same set of questionnaires administered to both groups, yielding a post-treatment measure for P2 and an extended follow-up point for P1.

The certified coach delivered all three planned coaching sessions following an identical sequence, structure, and set of materials across both project teams. A staggered schedule was implemented to maintain parity and minimize contamination: during the first intervention phase (T1–T2), the two subgroups of P1 (G1 and G2) received their sessions on consecutive days with identical content and facilitation. The same pattern was mirrored for P2 (G3 and G4) during its later intervention phase (T2–T3).
% Although both teams shared the same laboratory space, they were scheduled at distinct times to prevent cross-group discussion and diffusion of treatment effects.

During each session, a trained observer attended silently, completing a structured checklist to verify adherence to the protocol, timing, and sequence of activities. Any deviations from the planned procedure were immediately documented and discussed post-session to maintain consistency across groups. Observational notes were recorded digitally using a standardized template, capturing participant engagement, interaction quality, and contextual factors that could influence results.

Across all four measurement points, all participants completed the instruments, resulting in a full longitudinal dataset. Data integrity was continuously monitored by the SE research team to verify identifier consistency, response completeness, and timestamp accuracy. All materials used during operation (including the survey, coaching guides, anonymized dataset, consent form, and analysis scripts) are openly available in our repository \cite{repo}.

\subsection{Analysis Procedures}
Our analysis procedure consisted of three sequential stages: data preparation and validation, primary analysis of the main construct (CIPS), and secondary and integrative analyses incorporating the complementary scales and observational data. The procedure aimed to assess the direction and magnitude of change across measurement points while maintaining analytical transparency and coherence with the quasi-experimental design.

All questionnaire responses were checked for completeness, participant identification consistency, and chronological alignment across the four measurement points (T0–T3). No attrition occurred, and all participants completed the full set of instruments at each stage. Scores were computed following each instrument’s official manual (CIPS, WHO-5, SWLS, PANAS). Internal consistency was verified using Cronbach’s $\alpha$, which indicated acceptable reliability across all scales ($\alpha$ $\geq$ .78). The results indicated good reliability for the PANAS Positive ($\alpha$=0.85) and PANAS Negative ($\alpha$=0.88) subscales. The Satisfaction with Life Scale (SWLS) also demonstrated acceptable reliability ($\alpha$=0.73), exceeding the standard 0.70 threshold. The WHO-5 well-being scale showed lower internal consistency ($\alpha$=0.62); however, this value is often considered marginally acceptable for short (five-item) scales, suggesting that the instruments, overall, possess sufficient reliability for the present analysis.

% The distribution of scores was inspected visually (histograms and Q–Q plots)
% % and tested using Shapiro–Wilk, revealing non-normality in several measures; 
% and all inferential analyses relied on non-parametric statistics. 
To verify construct coherence, we calculated pairwise correlations among all scales at baseline, expecting negative relationships between CIPS and well-being measures (WHO-5, SWLS, PANAS-Positive) and positive correlations with distress (PANAS-Negative), consistent with prior evidence. This step confirmed the internal validity of the dataset before formal hypothesis testing.

The first analytical layer focused on evaluating whether the group coaching intervention produced measurable changes in impostor feelings, as assessed by the CIPS. Descriptive analyses (medians, interquartile ranges, and boxplots) were generated for each group (P1, P2) and each time point (T0–T3), allowing for visual inspection of distributional shifts and within-subject trajectories. The inferential comparison targeted the critical post-intervention point (T2), when P1 had completed coaching and P2 had not. Because of the small sample size (N=20) and non-normal data distribution, the Mann–Whitney U test was used to compare groups ($\alpha$ = .05). To assess practical relevance beyond statistical significance, effect sizes (Rosenthal’s r) and 95\% confidence intervals were computed. Within-group pre/post changes were also examined using Wilcoxon signed-rank tests (P1: T1→T2; P2: T2→T3) to evaluate the consistency of intervention effects. All analyses followed recommendations for quasi-experimental studies with low statistical power, emphasizing both direction and magnitude of change.

A secondary layer of analysis examined whether shifts in impostor feelings corresponded with concurrent changes in well-being (WHO-5), life satisfaction (SWLS), and affect (PANAS Positive/Negative). These analyses were primarily descriptive and visual, using boxplots to examine whether trends in psychological well-being paralleled the changes in CIPS across time. In addition, correlation matrices were computed at baseline (T0) and at the final follow-up (T3) for the aggregated sample, providing a multidimensional view of how the constructs interacted over the course of the intervention. These correlations also served to evaluate the theoretical coherence of the data.%, testing whether reductions in impostorism aligned with improvements in well-being and affective balance.

Finally, field notes from the non-participant observer were analyzed descriptively and organized chronologically to support data triangulation. Observation records were used to verify the fidelity of the intervention’s delivery and to identify contextual or behavioral factors that might have influenced participant responses. This integrative step was intended to strengthen conclusion validity by interpreting quantitative trends in light of team dynamics and environmental conditions observed during the sessions.

\section{Results}
\label{sec:experimentalresults}

%In this section, we present the experimental study results.

%\subsection{Study Results}

% This section details the study's results, beginning with the primary analysis of H1 (CIPS scores), then for H2 we present the comparative findings from the secondary psychological scales (WHO-5, SWLS, and PANAS), and report key noted from the non-participant observer.

This section presents the results, covering the H1 analysis (CIPS), the H2 comparative findings (WHO-5, SWLS, PANAS), and key non-participant observer notes.

\subsection{CIPS Results (H1)}

\begin{figure}[htbp]
\centerline{\includegraphics[width=\linewidth]{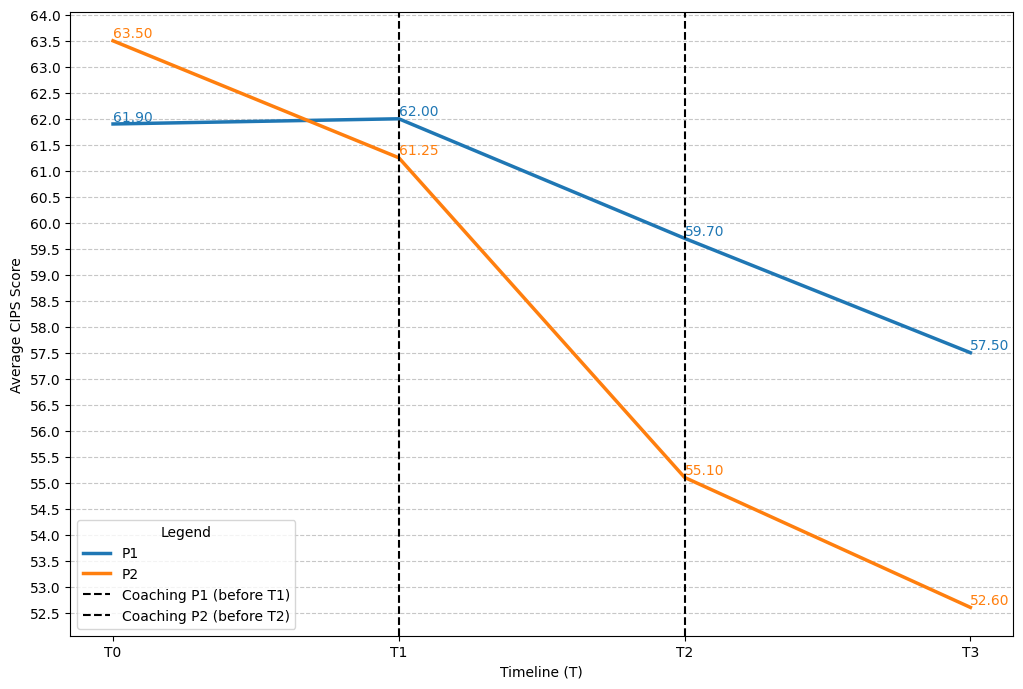}}
    \caption{Evolution of Average CIPS Scores throughout the experiment. The dashed vertical lines indicate the start of the coaching intervention for P1 (at T1) and P2 (at T2).}
    \Description{Line Graph with Average IP Score by Project}
    \label{fig:average_ip_score_by_t_and_project}
\end{figure}

Following the analysis procedures, we examined changes in IP using the CIPS across all four measurement points (T0–T3). Figure \ref{fig:average_ip_score_by_t_and_project} presents the evolution of CIPS scores for the initial treatment group (P1) and the control group (P2). During the pre-intervention period (T0 → T1), both groups remained relatively stable: P1’s average score was unchanged, while P2’s average decreased slightly.
% (from 63.50 to 61.25).

%As stated in our analysis procedures, we generated graphs to visualize the results from all survey scales, with a special attention to CIPS. These graphs provide a visual comparison of the score distributions for the initial treatment group (P1) and the control group (P2).

%We measured the baseline (T0) five weeks before the intervention started and the pre-intervention (T1) just before the intervention for the P1. Figure \ref{fig:average_ip_score_by_t_and_project} displays the CIPS average score distributions across all four time points (T0, T1, T2, T3). During the baseline and pre-intervention period (T0 to T1), both groups remained relatively stable; the median score for P1 was unchanged, while the median for P2 saw a slight decrease (from 63.50 to 61.25).

The first intervention phase (T1→T2) marked the critical comparison point. After receiving the coaching intervention, P1 exhibited a modest reduction in CIPS score, with its average score declining from 62.0 to 59.7 (-3.7\%). This decrease shifted the group’s average from the frequent IP range (61–80) to the upper boundary of the moderate IP range (41–60). However, during the same period, the control group (P2), which had not yet received any coaching, showed a sharper decline, from 61.25 to 55.10 (-10\%). In the second phase (T2→T3), when P2 received the same coaching intervention, its average dropped by an additional 4.5\%, while P1’s scores dropped only 3.6\%.

%The primary analysis occurred during the first intervention phase (T1 to T2). Following the coaching, the initial treatment group (P1) experienced a modest reduction, with their median CIPS score falling from 62 to 59.70 (a 3.8\% decrease). Notably, this change moved the group's median score from the 'frequent IP' range (61-80) to just below this cutoff, placing it within the 'moderate IP' range (scores 41–60). However, the most significant change during this period was observed in the control group (P2). Despite receiving no intervention, this group experienced a substantial drop in their median CIPS score, from 61.25 to 55.10, a reduction of approximately 10\%. Finally, in the second phase (T2 to T3), when P2 received the same coaching intervention, their median score decreased by an additional 4.5\%.

Moreover, a Mann–Whitney U test comparing both groups at T2 confirmed that these differences were not statistically significant ($U$ = 43.5, $p$ = .33). The corresponding Rosenthal effect size ($r$ = 0.10) also indicated negligible practical magnitude. Thus, while small downward shifts in CIPS were observed, the coaching intervention did not yield a measurable differential effect between the treatment and control groups within this timeframe.

%To test the primary hypothesis, a Mann-Whitney U test was performed to compare the CIPS scores at T2, after the first intervention. The results indicated no statistically significant difference between the two groups (U = 43.5, p = .33). Furthermore, the calculated Rosenthal's effect size was negligible (r = 0.10), suggesting the intervention had no meaningful impact on CIPS scores in P1 at T2 compared to the control project group.

% Regarding IP, as shown in Figure \ref{fig:average_ip_score_by_t_and_project}, CIPS score measured on T0 and T1 for P1 remained the same, whereas, for P2 sligtlhy went down (from 63,5 to 62). After the intervention for P1, in T2, the median CIPS score in the treatment group fell from 62 (at T1) to 59,7 (at T2), a reduction of 3,8\%. This indicates a mudança de faixa considerando que a nota de corte para dizer que a pessoa sofre de IP (ou seja, sofre de frequente a intensos sentimentos de impostor) é Score > 60. When P2 also received the treatment, T2-T3, houve uma queda de aproximadamente 4,5\%.

% However, when P2 was the control group, ou seja, wasnt receiving the coaching intervention, the Figure \ref{fig:average_ip_score_by_t_and_project} also depicts the greater decrease related so far, aproximately of 10\%.

\begin{figure}[htbp]
\centerline{\includegraphics[width=\linewidth]{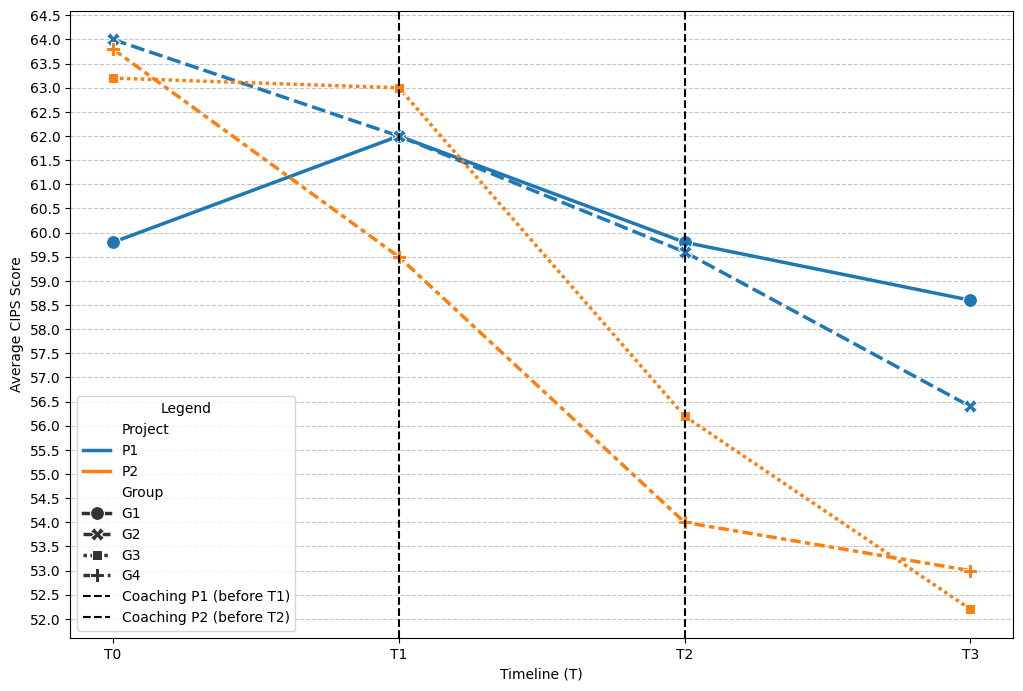}}
    \caption{Evolution of Average CIPS Scores by subgroups (G1-G4). This visualization confirms internal consistency within each project.}
    \Description{Line Graph with IP Score by Project and by Group}
    \vspace{-10pt} \label{fig:average_ip_score_by_group_com_grade}
\end{figure}

Figure \ref{fig:average_ip_score_by_group_com_grade} presents the CIPS trajectories for the four subgroups (G1–G4), illustrating internal consistency within each project. As described in our experiment operation, P1 comprised subgroups G1 and G2, while P2 comprised G3 and G4. During the first intervention phase (T1→T2), both treatment subgroups in P1 moved in close parallel: G1’s average CIPS score declined from 62.0 to 59.8 (–3.5\%), and G2’s from 62.0 to 59.6 (–3.9\%). The control project (P2) exhibited a similar internal coherence during its sharper decline in the same period, with G3 decreasing from 63.0 to 56.2 (–10.8\%) and G4 from 59.5 to 54.0 (–9.2\%). The parallel trends observed within both projects indicate that changes were systematic at the team level rather than driven by outlier behavior.

%Figure \ref{fig:average_ip_score_by_group_com_grade} depicts the average CIPS scores for the individual sub-groups (G1, G2, G3, G4). As noted in our methodology, P1 consisted of G1 and G2, while P2 consisted of G3 and G4. The graph demonstrates an internal consistency within each project during the first intervention phase (T1-T2). During this phase, the treatment sub-groups (P1) moved in parallel: G1's average score fell from 62.0 to 59.8 (a 3.5\% decrease), while G2's average fell similarly from 62.0 to 59.6 (a 3.8\% decrease).

%More notably, the control group (P2) showed the same internal consistency during its steep decline. In the same T1-T2 period, G3's average score dropped from 63.0 to 56.2 (a 10.8\% decrease), and G4's average score dropped from 59.5 to 54.0 (a 9.24\% decrease). This homogeneity suggests the effects observed in each project were systemic and not driven by a single sub-group.

\subsection{Multidimensional Results (H2)}
\begin{figure*}[htb!]
    \centering % Centraliza todo o conteúdo da figura

    % --- Todas as 5 imagens na mesma linha ---
    \begin{subfigure}[b]{0.19\textwidth} % Largura ajustada para caber 5
        \centering
        \includegraphics[width=\linewidth]{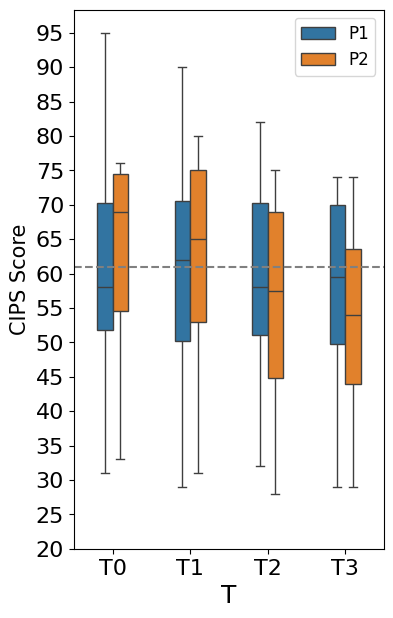}
        \caption{CIPS Score}
        \label{fig:box_plot_ip_scores_by_project_t}
    \end{subfigure}
    \hfill % Espaço flexível
    \begin{subfigure}[b]{0.19\textwidth} % Largura ajustada para caber 5
        \centering
        \includegraphics[width=\linewidth]{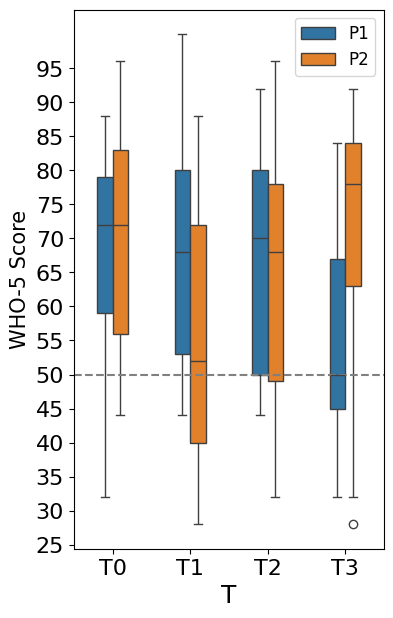}
        \caption{WHO-5 Score}
        \label{fig:box_plot_who_5_score_by_project_t}
    \end{subfigure}
    \hfill % Espaço flexível
    \begin{subfigure}[b]{0.19\textwidth} % Largura ajustada para caber 5
        \centering
        \includegraphics[width=\linewidth]{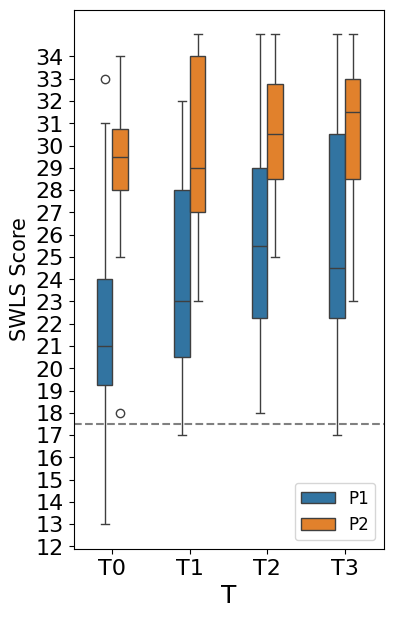}
        \caption{SWLS Score}
        \label{fig:box_plot_swls_score_by_project_t}
    \end{subfigure}
    \hfill % Espaço flexível
    \begin{subfigure}[b]{0.19\textwidth} % Largura ajustada para caber 5
        \centering
        \includegraphics[width=\linewidth]{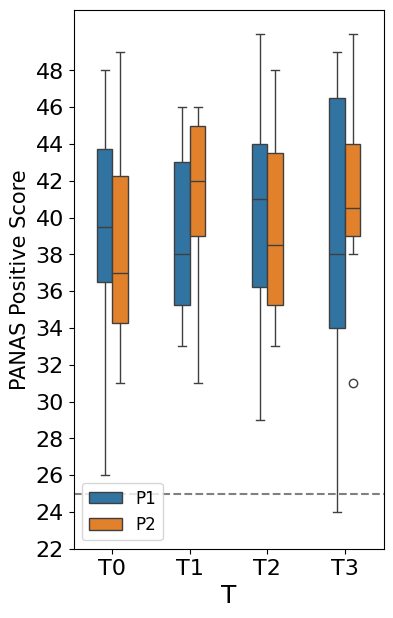}
        \caption{PANAS Positive Score}
        \label{fig:box_plot_panas_positive_score_by_project_t}
    \end{subfigure}
    \hfill % Espaço flexível
    \begin{subfigure}[b]{0.19\textwidth} % Largura ajustada para caber 5
        \centering
        \includegraphics[width=\linewidth]{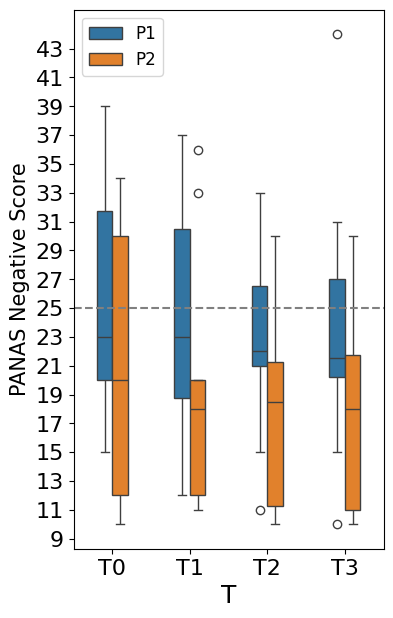}
        \caption{PANAS Negative Score}
        \label{fig:box_plot_panas_negative_by_project_t}
    \end{subfigure}

    \caption{Comparison of score distributions across the study timeline (T0–T3) for all five psychological scales. Dashed horizontal lines indicate clinical or categorical cut-offs for each instrument. The consistent parallel trends within each project (G1/G2 for P1 and G3/G4 for P2) indicate that the observed changes are systematic at the team level.}
    \label{fig:box_plots}
\end{figure*}

To contextualize the CIPS findings, we conducted a complementary analysis of the associated psychological constructs: subjective well-being (WHO-5), life satisfaction (SWLS), and affective balance (PANAS-Positive and PANAS-Negative). Figure \ref{fig:box_plots} presents the distributions of all measures across the data collection points (T0–T3).

%We now turn to the box plot analysis for all measured scales (CIPS, WHO-5, SWLS, PANAS-Positive, and PANAS-Negative) in order to increse the robustness of our exploratory analysis.

% To evaluate the effectiveness of a coaching intervention in mitigating the Impostor Phenomenon (CIPS) and its correlated variables (WHO5, SWLS, PANAS), a longitudinal study with a staggered control design was conducted. The treatment group (P1) received coaching between T1 and T2, while the control group (P2) received the same intervention in the subsequent phase, between T2 and T3.

At baseline (T0), P2 exhibited substantially higher CIPS score than P1, with median of 69.0 and 58.0, respectively. Despite this, P2 also reported greater life satisfaction (SWLS = 29.5 vs. 21.0 in P1), while both groups displayed identical well-being medians (WHO-5 = 72.0). The initial control period (T0→T1) revealed early volatility, most notably a sharp drop in P2’s well-being (WHO-5 decreasing from 72.0 to 52.0).

During the first intervention phase (T1→T2), when P1 received coaching, its median CIPS score slightly decreased from 62.0 to 58.0. However, the wait-list control (P2) experienced an even stronger spontaneous reduction (from 65.0 to 57.5) accompanied by a recovery in well-being (WHO-5 = 52.0 → 68.0). In the subsequent phase (T2→T3), when P2 underwent the same intervention, its positive trajectory continued: CIPS score declined further to 54.0, and well-being peaked at 78.0, the highest level observed across the study. In contrast, P1, now in follow-up, exhibited regression: CIPS scores rose modestly to 59.5, and well-being fell sharply from 70.0 to 50.0. Consequently, although both groups started with identical well-being (median = 72.0), they diverged markedly by T3, P2 improving to 78.0 while P1 declined to 50.0. Across the full period (T0→T3), P2 showed the most pronounced overall improvement, with CIPS score decreasing by 15 points (69.0 → 54.0) and concurrent increases in well-being and positive affect (PANAS-Pos).

\begin{figure}[htb!]
    \centering
        \begin{subfigure}[b]{.49\columnwidth}
            \centering
            \includegraphics[width=\linewidth]{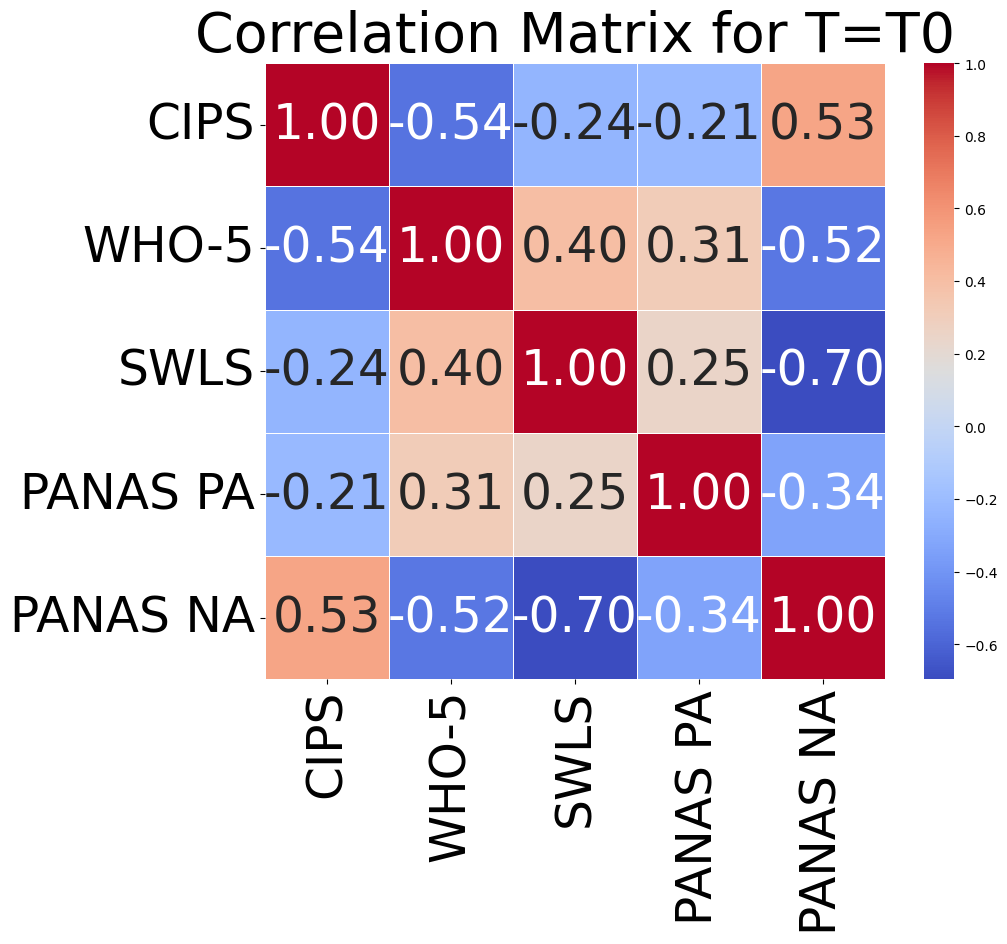}
            % \caption{Time T0}
            \label{fig:correlation_t0}
        \end{subfigure}
        % \hfill
        \begin{subfigure}[b]{.49\columnwidth}
            \centering
            \includegraphics[width=\linewidth]{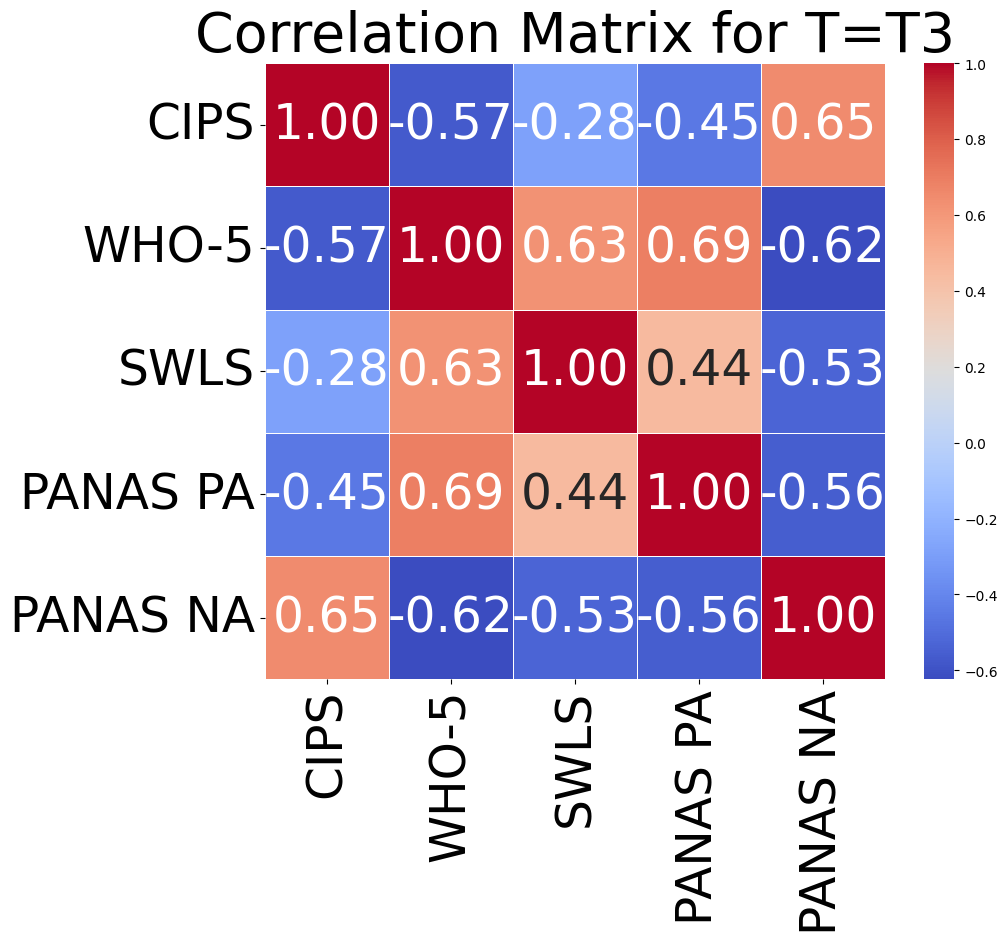}
            % \caption{WHO-5}
            \label{fig:correlation_t3}
        \end{subfigure}

    \caption{Correlation matrices of psychological scales at baseline (T0) and final follow-up (T3). The heatmaps illustrate the intensification of theoretically consistent relationships between impostor feelings (CIPS), well-being (WHO-5), and affect (PANAS) over the course of the study.}
    % \Description{Line Graph with IP Score by Profile}
    \vspace{-10pt} \label{fig:correlation_t0_t3}
\end{figure}

To further explore construct relationships, we compared the correlation matrices for the aggregated sample at baseline (T0) and final follow-up (T3) (see Figure \ref{fig:correlation_t0_t3}). At T0, CIPS displayed the expected pattern: a positive association with negative affect (PANAS-Neg, $\rho$ = 0.53) and a negative association with well-being (WHO-5, $\rho$ = -0.54). Among the supportive measures, the strongest relationship was between negative affect and life satisfaction ($\rho$ = -0.70), while the link between well-being and positive affect was weak ($\rho$ = 0.31).

%We performed a comparative analysis (figure \ref{fig:correlation_t0_t3} of the correlation matrices for the aggregated sample at baseline (T0) and at the final follow-up (T3).
%At the baseline, the CIPS Score showed the expected moderate correlations: it was positively associated with PANAS Negative (0.53) and negatively associated with WHO-5 (-0.50). Among the support measures, the strongest correlation was between PANAS Negative and SWLS (-0.70) , while the correlation between WHO-5 and PANAS Positive was surprisingly weak (0.28).

By T3, these relationships had intensified in theoretically consistent directions. The associations between impostor feelings and distress indicators strengthened (CIPS-PANAS-Neg = 0.65; CIPS-WHO-5 = -0.57), and the relationship between well-being and positive affect rose sharply from weak to very strong ($\rho$ = 0.69). These evolving correlations suggest that, over time, the constructs converged toward a more coherent affective pattern, with improvements in well-being closely tracking reductions in impostorism and negative affect.

%By T3, the relationships between the CIPS Score and distress indicators had strengthened (e.g., vs. PANAS Negative rose to 0.65; vs. WHO-5 strengthened to -0.60). The most drastic change occurred in the well-being measures: the correlation between WHO-5 (well-being) and PANAS Positive (positive affect) jumped from weak (0.28) to very strong (0.76).

% \begin{figure}[htbp]
% \centerline{\includegraphics[width=\linewidth]{Images/correlation_t0_t3.png}}
%     \caption{IP Score by Profile}
%     \Description{Line Graph with IP Score by Profile}
%     \label{fig:correlation_t0_t3}
% \end{figure}

In addition, observational data were collected to ensure intervention fidelity and to capture contextual nuances that could inform interpretation of the quantitative results. The non-participant observer confirmed that the certified coach delivered the protocol consistently across all sessions, following identical structures, materials, and instructions. While the facilitation was standardized, participants’ engagement varied. We noticed some preferred brief reflection before sharing, whereas others contributed spontaneously, creating a fluid and collaborative atmosphere that supported collective learning.

%Data from the non-participant observer was collected to ensure intervention fidelity and document contextual dynamics. The observer confirmed that the coach consistently delivered the same structured protocol to all sub-groups, including standardized introductions, confidentiality agreements, and task instructions. The coach's protocol was consistent, but the participant response style was flexible: for some questions, participants were given time to reflect individually before answering, while for others, the group responded spontaneously.

Adherence to home assignments was mixed. The documentary-based task, for example, saw limited completion, but the feedback-seeking exercise was almost universally performed. Members of P2 reported that such feedback routines were already part of their team practice, indicating a pre-existing culture of openness that may have amplified the intervention’s effects on well-being and communication. Across sessions, participants showed growing comfort with vulnerability and self-disclosure, particularly when discussing shared doubts and performance pressure.

%Home assignments were given at the end of each session, but adherence was inconsistent. For instance, a task requiring participants to watch a specific documentary was not completed by all members. In contrast, a task requiring participants to request feedback from three people nearly all participants completed. Moreover, when this topic was discussed in-session, participants from Project 2 volunteered that they had already implemented their own team-native feedback dynamic and were familiar with its benefits.

In the closing sessions, participants articulated distinct takeaways. Members of P1 emphasized normalization and self-awareness, describing relief in recognizing shared experiences (“\textit{It’s good to see that sometimes I’m not crazy}”) and gains in confidence (“\textit{It was important to realize things about myself, to risk more, [and] speak up more}”). They also valued the collective nature of the process: “\textit{What I liked most was doing this in a group, there are similar things, different things, [and] I learned a little more}”. On the other hand, participants from P2 focused on self-compassion and interpersonal learning, expressing a relational framing of growth: “\textit{[We don’t] need to keep blaming or punishing [ourselves] for things that are not in [our] control}”, and “\textit{Asking for feedback seems to be very productive instead of just waiting passively}”. They also linked these discussions to enhanced “\textit{communication}” and “\textit{empathy}” within their teams.

%These results align with the quantitative findings: P1’s gains centered on cognitive reframing, while P2’s improvements in well-being appeared tied to relational and emotional mechanisms. In summary, the qualitative evidence reinforces that group coaching modulated impostor feelings through both individual reflection and collective emotional regulation.

%During the final session, participants from both projects reported their key insights. Participant from P1 shared insights highlighting the value of normalizing their feelings ("It's good to see that sometimes I'm not crazy") and the intervention's impact on self-awareness and behavior ("It was important to realize things about myself, to risk more, [and] speak up more"). This participant also endorsed the group modality: "What I liked most was doing this in a group; there are similar things, different things, [and] I learned a little more."

%Verbal data from  P2 included statements on self-compassion ("[We don't] need to keep blaming or punishing [ourselves] for things that are not in [our] control") and the value of proactive feedback ("Asking for feedback seems to be very productive instead of just waiting passively"). Participants from this group also noted that the discussions helped improve "communication" and "empathy."

\section{Discussion}
\label{sec:discussion}

%In this section we discuss the obtained results.

%\subsection{Interpretation of Findings}

% comparar o q foi feito nesse artigo e o q achamos e oq o outro artigo achou

% falar sobre o resultado do restante das escalas

% contexto do projeto 2 de dinamica / feedback

 % [discussao]Although not significant, the direction of the averages suggests that the IP of participants of both projects tended to reduce throughout the quasi-experiment timeline. Com coaching reduziu, mas teve outros fatores que também podem ter influenciado a redução, em particular ... (CNV)

The results provide only partial support for our hypotheses. Concerning H1, while participation in the group-coaching intervention was associated with a slight reduction in CIPS scores among the treatment group (P1), the most pronounced decrease occurred unexpectedly in the control group (P2) during the same observation period. When P2 subsequently received the same intervention, their additional improvement was minimal. Likewise, regarding H2, no consistent gains were observed across the well-being (WHO-5), life satisfaction (SWLS), or affect (PANAS) scales that could be attributed to the intervention itself. These results suggest that the observed changes were potentially more closely linked to contextual, social, or temporal factors embedded in the working environment than to the structured coaching sessions alone.

%The results indicate that the coaching intervention produced limited quantitative effects on impostor feelings. While the treatment group (P1) showed a modest reduction in CIPS scores following the coaching sessions, the most pronounced decrease occurred unexpectedly in the control group (P2) during the same period. When P2 later received the same intervention, their additional improvement was minimal. This result suggests that the observed reductions in CIPS scores were not primarily driven by the structured coaching itself, but eventually by contextual, social, and temporal factors embedded in the participants’ working environment.

%Our analysis shows that while the treatment group P1 experienced some reduction in their average IP score following the coaching intervention (as seen in Figure \ref{fig:average_ip_score_by_t_and_project}, between T1 and T2), the primary quantitative finding was the unexpected and more substantial drop in IP scores within the control group P2 during the same period. Furthermore, the intervention's limited impact was reinforced in the second phase: when P2 (now the treatment group) received the same coaching protocol (between T2 and T3), their subsequent reduction in IP scores was also low.

Importantly, the internal consistency observed across project sub-groups (see Figure \ref{fig:average_ip_score_by_group_com_grade}) reinforces that these effects were systemic rather than random. To interpret these findings, we integrate quantitative, qualitative, and contextual evidence to identify plausible mechanisms (spanning project complexity, team climate, and measurement reactivity) that jointly shaped participants’ trajectories found in our study.

%A deeper analysis of the project sub-groups (G1 and G2 in P1; G3 and G4 in P2) reinforces this finding by demonstrating internal consistency as seen in Figure \ref{fig:average_ip_score_by_group_com_grade}. The modest reduction (around 3\%) observed in P1 was seen similarly across both its sub-groups (G1 and G2). Likewise, the drop in P2 was consistently observed across its sub-groups (G3 and G4). This homogeneity between group projects suggests the effects were not random but rather a systemic characteristic of each project's unique environment.

%Therefore, an analysis of our quantitative data, contextualized by our qualitative data and the study setting, suggests several plausible, and potentially overlapping, explanations for this finding:

\textit{\textbf{Project Lifecycle and Complexity}}.
The four-month study period coincided with the full development lifecycle of both industry-linked projects, making temporal dynamics and task complexity central contextual factors. Consistent with Clance’s impostor cycle \cite{clance1985impostor}, impostor feelings often peak at the start of demanding tasks and subside temporarily upon successful performance. Both project groups displayed this trajectory: CIPS scores declined between the project’s midpoint and delivery, suggesting that progress and accomplishment may have naturally alleviated anxiety.

%Our analysis must account for the projects' temporal dynamics and inherent complexity as an explanatory variable. Our longitudinal measurement (T0-T3) was aligned with the four-month project development cycle. As shown in Figure 1, a comparison of the T0 and T3 endpoints reveals that reported impostor phenomenon scores indeed decreased for both projects. For instance, for P2, scores decreased by more than 10 points on the CIPS scale. This suggests a plausible link between IP fluctuations and the project's life-cycle stages (from initial planning to peak execution and final delivery). 

%This pattern can also be interpreted through the lens of the theoretical "impostor cycle." This model posits that the beginning of a new project or challenging task often induces heightened anxiety and fear of failure (which would align with the higher T0 scores). Upon the project's successful completion, the individual experiences a period of 'brief relief'. Therefore, the observed drop in IP at T3 may not reflect a permanent reduction, but rather this temporary 'relief' phase following the project's conclusion.

Differences in technical challenge further clarify the asymmetry between teams. In this regard, P1 tackled an automation problem in an unfamiliar domain requiring legacy-system integration, conditions known to heighten cognitive load and uncertainty. In contrast, P2 employed familiar technologies to build a predictive model with clear milestones. This discrepancy likely moderated how quickly team members experienced mastery and internalized competence. Thus, fluctuations in impostor feelings may partly reflect progress-based confidence rather than the direct impact of coaching. From a Social Cognitive Theory perspective \cite{bandura1986social}, the sense of efficacy derived from successful task performance could itself act as a stronger source of impostor reduction than external facilitation.

%The varying complexity across different projects likely moderated this effect. Consequently, the progression through the sprint cycle, with its associated stressors and milestones, presents a significant alternative or parallel explanation for the observed changes in IP scores. The P1, which exhibited the smallest decrease in IP, involved an unfamiliar domain, significant technological challenges, and integration with a legacy system. In contrast, P2 involved familiar technology and a predictive modeling task, where the primary challenge was data acquisition. It is plausible that by the T1-T2 measurements, the P2 team had already defined the project's core methodology, while the P1 team continued to face high uncertainty.

\textit{\textbf{Pre-Existing Team Dynamics and Psychological Safety}}. Qualitative data revealed that P2 team had already institutionalized reflective and empathic communication practices before the intervention. During observations, members described an internally developed feedback routine aligned with principles of Nonviolent Communication (NVC) \cite{adriani2024non}, which emphasizes empathy, self-expression without judgment, and constructive feedback. These environments foster psychological safety (a shared belief that taking interpersonal risk is accepted and valued \cite{edmondson1999psychological}). Prior studies show that teams with higher psychological safety exhibit stronger self-efficacy and lower social evaluation anxiety \cite{alami2023antecedents, cerqueira2024empathy}.

%As a quasi-experiment conducted in a real-world organizational setting, the study was subject to uncontrolled variables inherent to dynamic work environments. For instance, during the coaching sessions administered to team P2 after the T2 measurement, the observer noted that the group discussed their internal practices. They reported having autonomously implemented a new, structured feedback dynamic. They reported they had independently implemented their own structured feedback dynamic.
% called "Speedwraps". Speedwraps \cite{Marinho2017Speedwrap} 
%Furthermore, this dynamic was designed to create a safe environment for team members to exchange simple, sincere, and honest horizontal feedback. It adheres to the principles of Non-Violent Communication (NVC), favoring effective communication through empathy and supporting the expression of feelings and needs rather than critiques or value judgments. Feedback was exchanged in pairs, with each conversation held privately between the two participants. This dynamic was not present in P1.

The spontaneous adoption of these practices may explain the substantial decrease in impostor scores observed in P2 even before receiving coaching. By reducing fear of judgment and promoting self-acceptance, the team’s climate reproduced some of the mechanisms targeted by formal coaching, including reframing negative self-attributions and building emotional stability. Indeed, participants from the P2 team reported that the discussions helped improve 'communication' and 'empathy.', which is a crucial skill for software practitioners, supporting them in building better products and improving their work environment \cite{cerqueira2024empathy}. In this sense, the informal peer feedback process may have achieved similar psychological outcomes to the structured intervention, highlighting how embedded social practices can catalyze cognitive and affective change more effectively than time-bounded sessions.

\textit{\textbf{Diffusion of Treatment and Reactivity Effects}}. Because both project teams worked in the same physical space, diffusion of treatment could be an impactful issue. Participants from P1 (the initial treatment group) likely discussed session topics and reflective exercises with P2 members, who may have benefited from these insights before their own intervention phase. Such contamination, well documented in quasi-experimental field studies \cite{wohlin2024experimentation}, may have potentially impacted between-group differences.

Another plausible factor is measurement reactivity \cite{campbell1968quasi}. Completing repeated self-assessment scales (CIPS, WHO-5, SWLS, and PANAS) required participants to engage in structured introspection about their competence, emotions, and life satisfaction. This repeated reflection may itself function as a micro-intervention by heightening self-awareness and prompting cognitive reframing. Notably, the control group’s largest improvement occurred immediately after the pre-tests, aligning with this interpretation. The combination of survey reactivity and cross-group dialogue may therefore explain much of the unexpected reduction in IP during the initial observation phase.

\textit{\textbf{Implications for SE Research and Practice}}. The findings offer theoretical interesting implications into how the IP operates within software engineering contexts. Rather than a fixed personality trait, impostor feelings emerged as fluid and socially mediated experiences shaped by project complexity, temporal dynamics, and team climate. The observed fluctuations in CIPS scores suggest that progress and collective reflection (rather than formal coaching) were also drivers of change. Thus, the stronger improvement among teams that had already cultivated empathic feedback practices may underscores the role of psychological safety \cite{edmondson1999psychological} as a moderating mechanism. These results refine existing models of the impostor cycle \cite{clance1985impostor} by situating it within the socio-technical realities of collaborative software work.

%Practically, our findings suggest that reducing impostor feelings among early-career software engineers may rely less on standalone interventions and more on cultivating reflective and psychologically safe work environments. Rather than treating IP as an individual deficit to be “coached away”, organizations could embed brief, structured opportunities for reflection and feedback into routine practices. When integrated naturally into team workflows, these moments can normalize vulnerability, foster mutual recognition of uncertainty, and transform routine technical discussions into spaces that also sustain emotional awareness and collective learning.

In practice, our findings suggest that mitigating impostor feelings among early-career software engineers may depend less on isolated coaching initiatives and more on fostering reflective and feedback-oriented team environments. In practical actions, teams may take into account encouraging brief, structured moments of self-assessment and dialogue within existing practices (such as retrospectives, code reviews, or mentoring interactions) could help normalize doubt and cultivate openness. Rather than positioning coaching as a standalone intervention, organizations would consider on integrating its reflective principles into everyday collaboration, promoting psychological safety as an ongoing and shared responsibility within teams.

%Practically, the findings from this work suggest that mitigating impostor feelings among early-career software engineers may also benefit from fostering reflective and feedback-oriented team environments. Encouraging brief, structured moments of self-assessment and dialogue within existing team practices—such as retrospectives, code reviews, or mentoring conversations—could help normalize experiences of doubt and promote a climate of openness. Rather than positioning coaching as a standalone solution, future organizational efforts might focus on integrating its reflective principles into everyday collaboration.

%Practically, the study suggests that mitigating impostor feelings among early-career software engineers may depend less on isolated coaching programs and more on embedding reflective and feedback-oriented practices into everyday work routines. Integrating short and structured moments of self-assessment and dialogue into Agile ceremonies, code reviews, or mentoring interactions may normalize vulnerability and cultivate a climate of openness. 

% Educational programs and organizations can further support this by fostering peer feedback structures (such as mentoring pairs or NVC inspired sessions) that build self-efficacy and emotional resilience over time. Therefore, addressing IP in SE therefore requires designing social and organizational systems that allow confidence, competence, and belonging to develop together.

\section{Threats To Validity}
\label{sec:threats}

%In this session, we describe the threats to validity using the four-part classification by Wohlin et al. \cite{wohlin2024experimentation} (construct, internal, external, and conclusion). 

Regarding the \textbf{Internal Validity}, we acknowledge the absence of random assignment may introduce potential selection bias, as the two project teams differed in task complexity, technological domain, and project maturity. The effects of maturation may also have influenced the outcomes, given the four-month period during which the participants naturally gained confidence and skills through project work. Contextual factors further shaped results since one team had independently implemented a feedback practice based on Nonviolent Communication, which likely enhanced self-awareness and psychological safety. Diffusion of treatment was another plausible risk, since both teams shared the same workspace and could informally exchange reflections about the sessions. Repeated exposure to the same self-assessment instruments may also have produced testing effects, whereby reflection on questionnaire items prompted cognitive or emotional changes independent of coaching.  Sample size also limits the ability to attribute changes in IP directly to the coaching intervention. To mitigate these issues, we employed a wait-list design to ensure both groups eventually received the intervention, balanced group sizes to reduce structural inequity, and used a consistent facilitation protocol to minimize procedural variance.

Concerning \textbf{External Validity}, the study’s generalizability is constrained by its small and homogeneous sample of early-career software engineers engaged in academic–industry projects within an university innovation lab. Although these projects mirrored professional practice through agile routines and client interaction, the academic setting offered mentorship and psychological safety uncommon in most workplaces. These contextual characteristics, together with the limited sample size, restrict inference to similar early-career or educational environments. Accordingly, our results should be interpreted as exploratory and generative.

Threats to \textbf{Construct Validity} were mitigated through the use of validated psychometric instruments (CIPS, WHO-5, SWLS, PANAS) and a standardized coaching protocol verified by a non-participant observer. The same certified coach facilitated all sessions using identical materials and sequencing to ensure treatment fidelity. Participants were blinded to the study’s focus on the Impostor Phenomenon and informed only that the activity related to general well-being, reducing demand characteristics and social desirability bias. Nonetheless, the reliance on self-report measures introduces common-method variance. Future research could strengthen construct validity by triangulating quantitative scales with behavioral, peer, or physiological data.

%Construct validity refers to the degree to which our study measures and manipulates the intended theoretical constructs. The main threat regarding construct validity is using improper instruments. We address threats in different ways.
%First, to ensure we were accurately measuring our dependent variables, we used evaluated instruments (CIPS, WHO-5, PANAS, and SWLS), which are widely accepted in the literature and research.

%Second, to address the construct validity of the intervention (the "cause"), we employed a structured protocol and an external observer to ensure the fidelity of the intervention, verifying that the same protocol was delivered consistently to both projects.

%Finally, a significant threat in intervention studies is participant reactivity (such as social desirability bias or demand characteristics), where participants' responses are influenced by their knowledge of the study's goals. We mitigated this threat by blinding the participants to the primary hypothesis. Participants were informed they would receive a general "well-being training" inherent to their lab activities, but they were unaware that the study's focus was on the Impostor Phenomenon. This blinding minimizes the risk that participants would alter their self-reported IP scores to match perceived researcher expectations.

Finally, in terms of \textbf{Conclusion Validity}, the small sample size inherently limited statistical power, heightening the risk of Type II errors. To address this, we used non-parametric tests (Mann–Whitney U) and reported effect sizes to contextualize the direction and magnitude of observed changes. Researcher expectancy also represents a potential bias, as one author served as the non-participant observer. This issue was mitigated through structured observation templates and post-session debriefing with the psychology coauthors to discuss interpretations. Although statistical significance was not achieved, reporting both descriptive and effect-size indicators supports transparency and provides an empirical foundation for future investigation \cite{wohlin2024experimentation, cumming2014new, lakens2013calculating}.
\section{Concluding Remarks}
\label{sec:conclusion}

This novel study offered an exploratory and quasi-experimental evaluation of a group coaching intervention designed to mitigate impostor feelings among early-career software engineers. Using a wait-list control design, we observed modest reductions in impostor scores across both project teams, though the most pronounced improvement occurred in the control group during the observation period. These findings suggest that impostor experiences are influenced by a combination of contextual and team-level factors (such as project complexity, feedback culture, and developmental stage) rather than by structured interventions alone.

Our contribution lies in documenting this complexity through a combination of quantitative and observational data. By situating a psychological intervention within a realistic and project-based environment, the study offers practical findings into how impostor feelings fluctuate in tandem with real work dynamics rather than in isolation. While the results do not confirm the efficacy of coaching as a primary mitigation strategy, they also reveal the importance of examining impostor experiences through an environmental lens that includes organizational and temporal dimensions.

Future research could examine how team environments influence impostor feelings. The control group’s improvement suggests that team-level factors, such as feedback culture and communication dynamics, can play a larger role than formal coaching. Studies using longitudinal or ethnographic methods could explore how teams naturally develop psychological safety, manage project stress, and build resilience through practices like Nonviolent Communication. 

%Future research should build upon this environmental perspective. A primary implication of our study is that the control group's significant improvement points to the power of emergent, team-level factors. Rather than focusing solely on formal, top-down interventions like coaching, we call for future studies to investigate how teams organically cultivate psychological safety and supportive dynamics. Longitudinal, observational research (perhaps using ethnographic methods) is needed to understand how teams autonomously manage feedback (e.g., through frameworks like Nonviolent Communication - NVC), navigate project-life-cycle stressors, and build resilience against impostor feelings. Understanding these naturally occurring mechanisms will be crucial for developing sustainable strategies that genuinely support software engineers.

% Future research should build upon these findings by testing varied intervention modalities (such as individual coaching or team-based feedback programs) across larger and more diverse samples. Replications in industrial-based and diverse settings could clarify which mechanisms most effectively support psychological safety and confidence among early-career engineers.

\section*{Artifacts Availability}
To promote transparency and reproducibility, all study materials are openly available in our public repository \cite{repo}.

  \begin{acks}

We express our gratitude to the Brazilian Research Council - CNPq (Grant 312275/2023-4), Rio de Janeiro State's Research Agency - FAPERJ (Grant E-26/204.256/2024), the Coordination for the Improvement of Higher Education Personnel (CAPES), the  Kunumi Institute, and University of Bari TNE-DeSK project “Transnational Education Initiatives-Developing Shared Knowledge
in Smart Materials and Digital Transition for Sustainable Economy”, for their generous support.
\end{acks}
\bibliographystyle{ACM-Reference-Format}
\bibliography{sample-base}

% %%
% %% If your work has an appendix, this is the place to put it.
% \appendix









\end{document}